\documentclass[12pt]{jpconf}

\usepackage{graphicx}
\usepackage{subfig}
\usepackage{dcolumn}
\usepackage{bm}
\usepackage{enumerate}
\usepackage{subfig}
\usepackage{rotating}
\usepackage{multirow}
\usepackage{psfrag}
\usepackage{gensymb}
\usepackage{amsfonts}
\usepackage{amssymb}
\usepackage{latexsym}
\newcommand{\newc}{\newcommand}
\newc{\ra}{\rightarrow}
\newc{\lra}{\leftrightarrow}
\newc{\beq}{\begin{equation}}
\newc{\eeq}{\end{equation}}
\newc{\barr}{\begin{eqnarray}}
\newc{\earr}{\end{eqnarray}}

\def\barr{\begin{eqnarray}}
\def\earr{\end{eqnarray}}

\begin{document}

\title[direct WIMP searches]{ Predicted rates for direct WIMP searches}

\author{ J. D. Vergados}

\address{ Theoretical Physics Division, University
of Ioannina, Ioannina, Gr 451 10, Greece}
\ead{vergados@uoi.gr}
 \begin{abstract}
The differential event rate for direct detection of dark matter, both the time averaged and the modulated one due to the motion of the Earth, are discussed. The calculations focus on relatively light cold dark matter candidates  (WIMP) and low energy transfers. It is shown that  for some WIMP masses the modulation amplitude may change sign. This effect can be exploited to yield information about the mass of the dark matter candidate.
\end{abstract}

\section{Introduction}
The combined MAXIMA-1 \cite{MAXIMA-1}, BOOMERANG \cite{BOOMERANG},
DASI \cite{DASI} and COBE/DMR Cosmic Microwave Background (CMB)
observations \cite{COBE} imply that the Universe is flat
\cite{flat01}
and that most of the matter in
the Universe is Dark \cite{SPERGEL,WMAP06},  i.e. exotic. Combining 
the data of these quite precise experiments one finds:
$$\Omega_b=0.0456 \pm 0.0015, \quad \Omega _{\mbox{{\tiny CDM}}}=0.228 \pm 0.013 , \quad \Omega_{\Lambda}= 0.726 \pm 0.015.$$
Since any "invisible" non exotic component cannot possibly exceed $40\%$ of the above $ \Omega _{\mbox{{\tiny CDM}}}$
~\cite {Benne}, exotic (non baryonic) matter is required and there is room for cold dark matter candidates or WIMPs (Weakly Interacting Massive Particles).

Even though there exists firm indirect evidence for a halo of dark matter
in galaxies from the
observed rotational curves, see e.g the review \cite{UK01}, it is essential to directly
detect
such matter.
The possibility of such detection, however, depends on the nature of the dark matter
constituents and their interactions.

Since the WIMP's are  expected to be
extremely non relativistic, with average kinetic energy $\langle T\rangle  \approx
50 \ {\rm keV} (m_{\mbox{{\tiny WIMP}}}/ 100 \ {\rm GeV} )$, they are not likely to excite the nucleus.
So they can be directly detected mainly via the recoiling of a nucleus
(A,Z) in elastic scattering. The event rate for such a process can
be computed from the following ingredients\cite{JDV12a} :
i) The elementary nucleon cross section.
ii) knowledge of the relevant nuclear matrix elements
\cite{Ress,DIVA00}, obtained with as reliable as possible many
body nuclear wave functions, iii) knowledge of the WIMP density in our vicinity and its velocity distribution.

The nucleon cross sections  can also be extracted from the data of event rates, if and when such data become available. From limits on the event rates, one can obtain exclusion plots on the nucleon cross sections as  functions of the WIMP mass. 
In the standard nuclear recoil experiments, first proposed more than 30 years ago \cite{GOODWIT}, one has to face the problem that the reaction of interest does not have a characteristic feature to distinguish it
from the background. So for the expected low counting rates the background is
a formidable problem. Some special features of the WIMP-nuclear interaction can be exploited to reduce the background problems. Such are:

i) the modulation effect: this yields a periodic signal due to the motion of the earth around the sun. Unfortunately this effect, also proposed a long time ago \cite{Druck} and subsequently studied by many authors \cite{PSS88,GS93,RBERNABEI95,LS96,ABRIOLA98,HASENBALG98,JDV03,GREEN04,SFG06}, is small and becomes even smaller than  $2\%$ due to cancelations arising from nuclear physics effects,

ii) backward-forward asymmetry expected in directional experiments, i.e. experiments in which the direction of the recoiling nucleus is also observed. Such an asymmetry has also been predicted a long time ago \cite{SPERGEL88}, but it has not been exploited, since such experiments have been considered  very difficult to perform, but they now appear to be feasible\cite{JDV12a}. 
iii) transitions to excited states: in this case one need not measure nuclear recoils, but the de-excitation $\gamma$ rays. This can happen only in very special cases since the average WIMP energy is too low to excite the nucleus. It has, however, been found that in the special case of the target $^{127}$I such a process is feasible \cite{VQS04} with branching ratios around $5\%$,
(iv) detection of electrons produced during the WIMP-nucleus collision \cite{VE05,MVE05} and
v) detection of hard X-rays produced when the inner shell holes are filled\cite{MouVerE}.


  
In the present paper we will limit our attention to the standaerd recoil experiments and study the differential   event rates, both time averaged and modulated,  in the region of low energy transfers, as in the DAMA experiment \cite{DAMA1,DAMA11}, focusing our attention on relatively light WIMPS \cite{XENON10,CoGeNT11,FPSTV11}. Such light WIMPs  can be accommodated in some SUSY models \cite{CKWY11}. We will also present some results on the total rates as well.  We will employ here  the standard Maxwell-Boltzmann (M-B)
distribution for the WIMPs of our galaxy and we will not be concerned with other  distributions 
\cite{VEROW06,JDV09,TETRVER06,VSH08},
even though some of them
 may affect the modulation.  The latter will be studied elsewhere. We will explicitly show that the modulation amplitude, entering both the differential and the total rates, changes sign for certain reduced WIMP-nuclear masses. As a result such an effect, if and when the needed data become available, may be exploited to infer the WIMP mass.


\section{The formalism for the WIMP-nucleus differential event rate}
This formalism adopted in this work is well known (see e.g. the recent reviews \cite{JDV06a,VerMou11}). So we will briefly discuss its essential elements here.
The differential event rate can be cast in the form:
\beq
\frac{d R}{ d Q}|_A=\frac{dR_0}{dQ}|_A+\frac{d{\tilde H}}{dQ}|_A \cos{\alpha}
\eeq
where the first term represents the time averaged (non modulated) differential event rate, while the second  gives the time dependent (modulated) one due to the motion of the Earth (see below). Furthermore
\barr
\frac{d R_0}{ d Q}|_A&=&\frac{\rho_{\chi}}{m_{\chi}}\frac{m_t}{A m_p} \sigma_n\left (\frac{\mu_r}{\mu_p} \right )^2 \sqrt{<\upsilon^2>} A^2\frac{1}{Q_0(A)}\frac{d t}{du}\nonumber\\
\frac{d {\tilde H}}{ d Q}|_A&=&\frac{\rho_{\chi}}{m_{\chi}}\frac{m_t}{A m_p} \sigma_n\left (\frac{\mu_r}{\mu_p} \right )^2 \sqrt{<\upsilon^2>} A^2 \frac{1}{Q_0(A)} \frac{d h}{du}
\label{drdu}
\earr
with with $\mu_r$ ($\mu_p$) the WIMP-nucleus (nucleon) reduced mass, $A$ is the nuclear mass number and $\sigma_n$ is the elementary WIMP-nucleon cross section. $ m_{\chi}$ is the WIMP mass and $m_t$ the mass of the target. 
Furthermore one can show that
\beq
\frac{d t}{d u}=\sqrt{\frac{2}{3}} a^2 F^2(u)   \Psi_0(a \sqrt{u}),\quad \frac{d h}{d u}=\sqrt{\frac{2}{3}} a^2 F^2(u) \Psi_1(a \sqrt{u})
\eeq
with $a=(\sqrt{2} \mu_r b \upsilon_0)^{-1}$, $\upsilon_0$ the velocity of the sun around the center of the galaxy and $b$ the nuclear harmonic oscillator size parameter characterizing the nuclear wave function.  $ u$ is the energy transfer $Q$ in dimensionless units given by
\begin{equation}
 u=\frac{Q}{Q_0(A)}~~,~~Q_{0}(A)=[m_pAb^2]^{-1}=40A^{-4/3}\mbox{ MeV}
\label{defineu}
\end{equation}
and $F(u)$ is the nuclear form factor. Note that the parameter $a$ depends both on the WIMP , the target and the velocity distribution. Note also that for a given energy transfer $Q$ the quantity $u$ depends on $A$.\\
The functions $\Psi_0(a \sqrt{u})$ and $\Psi_1(a \sqrt{u})$ for a M-B distribution take the following form:
\beq
\Psi_0(x)=\frac{1}{2}
   (\mbox{erf}(1-x)+\mbox{erf}(x+1)+\mbox{\tiny{erfc}}(1-y_{\mbox{\tiny{esc}}})+\mbox{erfc}(y_{\mbox{\tiny{esc}}}+1)-2)
\eeq
\barr
\Psi_1(x)&=&\frac{1}{4} \delta 
   \left( -\mbox{erf}(1-x)-\mbox{erf}(x+1)-\mbox{erfc}(1-y_{\mbox{\tiny{esc}}})-
   \mbox{erfc}(y_{\mbox{\tiny{esc}}}+1) \right . \nonumber\\
  && \left . +\frac{2 e^{-(x-1)^2}}{\sqrt{\pi }}
   +\frac{2
   e^{-(x+1)^2}}{\sqrt{\pi }}-\frac{2 e^{-(y_{\mbox{\tiny{esc}}}-1)^2}}{\sqrt{\pi
   }}-\frac{2 e^{-(y_{\mbox{\tiny{esc}}}+1)^2}}{\sqrt{\pi }}+2\right)
\earr
where erf$(x)$ and erfc$(x)$ are the error function and its complement respectively,  $\delta\approx 0.135$ is the ratio of the velocity of the Earth around the sun to that of the sun around the center of the galaxy and $\alpha$ is the phase of the Earth ($\alpha=0$, around June 3nd).

Sometimes we will write the differential rate as:
\beq
\frac{d R}{ d Q}|_A=\frac{\rho_{\chi}}{m_{\chi}}\frac{m_t}{A m_p} \sigma_n \left ( \frac{\mu_r}{\mu_p} \right )^2 \sqrt{<\upsilon^2>} A^2 \frac{1}{Q_0(A)}\left(\frac{d t}{du}(1+ H(a \sqrt{u}) \cos{\alpha}\right )
\label{dhduH}
\eeq
In this formulation $H(a \sqrt{u}) $, the ratio of the modulated to the non modulated differential rate, gives the relative differential modulation amplitude.

\begin{figure}
\begin{center}
\rotatebox{90}{\hspace{0.0cm} $H(a \sqrt{u})\rightarrow$}
\includegraphics[height=.30\textheight]{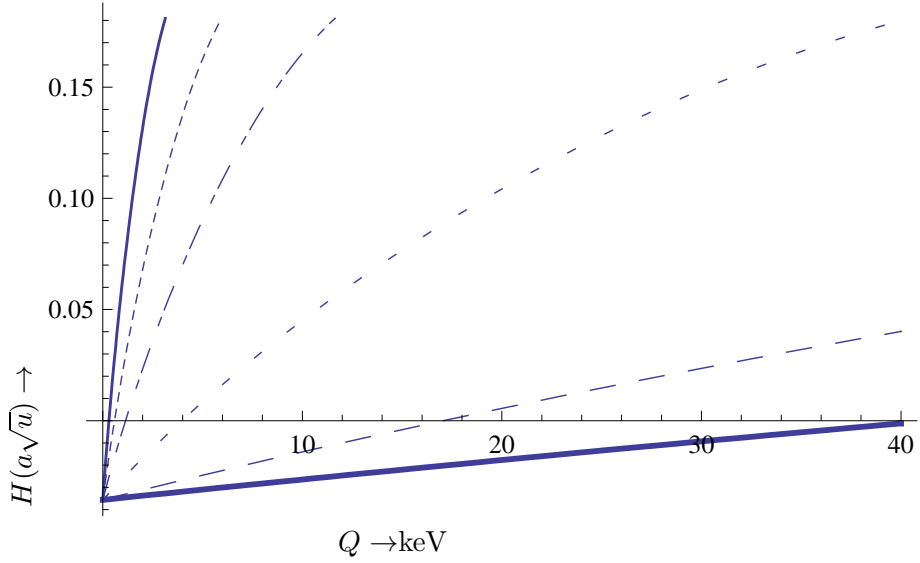}
\\
{\hspace{-2.0cm} $Q\rightarrow$keV}
\caption{ The  function $H(a \sqrt{u})$ entering   the modulated differential rate as a function of the recoil energy for a heavy target, e.g. $^{127}$I. Note that this is independent of  the form factor. The solid, dotted, dot-dashed, dashed, long dashed and thick solid lines correspond to 5, 7, 10, 20, 50 and 100 GeV WIMP masses.
 \label{fig:apsiH}}
\end{center}
\end{figure}

 The function $H(a \sqrt{u})$ is shown in Fig. \ref{fig:apsiH}. It is independent of the nuclear physics and  depends only on the reduced mass and the velocity distribution. They are thus the same for both the coherent and the spin mode. Note that $H(a \sqrt{u})$ can take both positive and negative values, which affects the location of the maximum.
\section{Some results on differential rates}
\begin{figure}
\begin{center}
\subfloat[]
{
\rotatebox{90}{\hspace{0.0cm} $dR/dQ\rightarrow$kg/(y keVee)}
\includegraphics[height=.17\textheight]{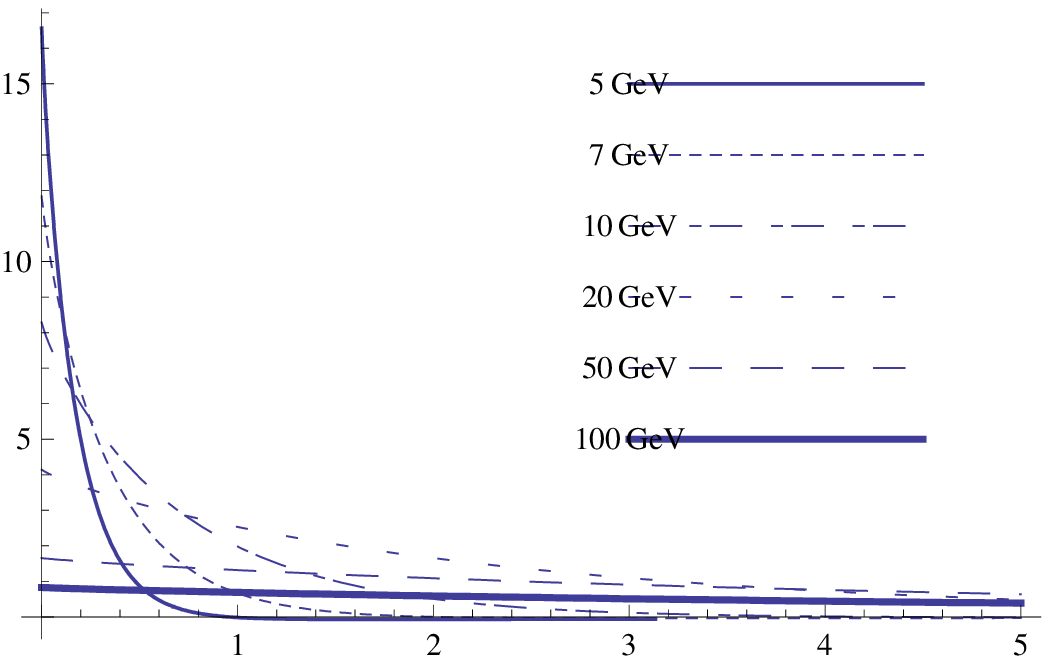}
}
\subfloat[]
{
\rotatebox{90}{\hspace{0.0cm} $d{\tilde H}/dQ\rightarrow$kg/(y keVee)}
\includegraphics[height=.17\textheight]{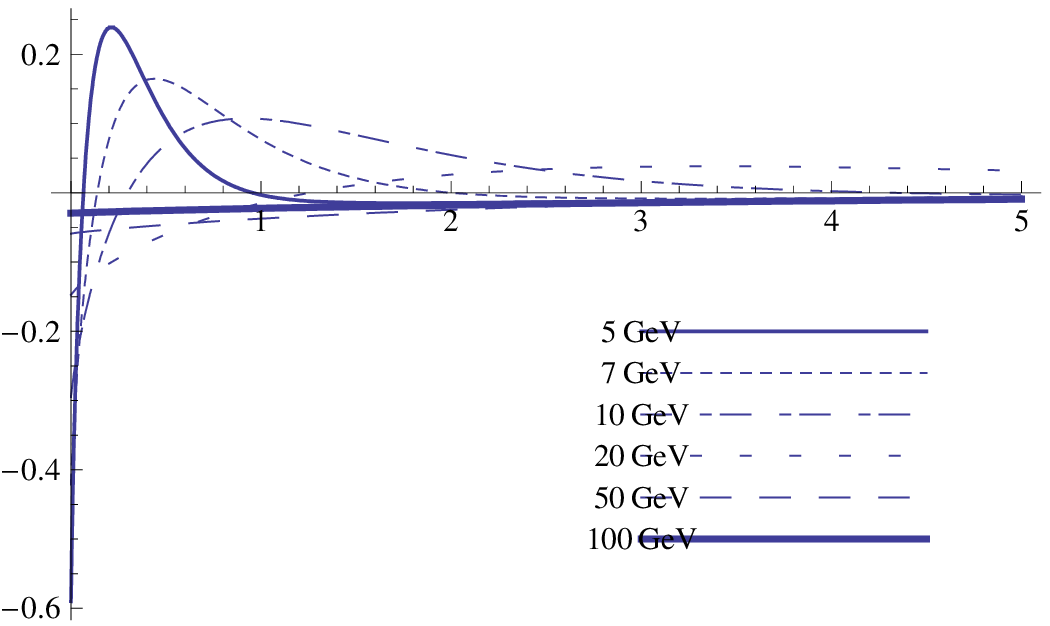}
}
\\
{\hspace{-2.0cm} $Q\rightarrow$keVee}
\caption{ The differential rate $\frac{dR}{dQ}$,   as a function of the recoil energy for a heavy target, e.g. $^{127}$I (a) and the amplitude for the modulated differential rate $\frac{d{\tilde H}}{dQ}$ (b), assuming a nucleon cross section of $10^{-7}$pb.  The solid, dotted, dot-dashed, dashed, long dashed and thick solid lines correspond to 5, 7, 10, 20, 50 and 100 GeV WIMP masses. Note that $\frac{d{\tilde H}}{dQ}$ is given in absolute units.
 \label{fig:dRdQdHdQ_127}}
\end{center}
\end{figure}

\begin{figure}
\begin{center}
\subfloat[]
{
\rotatebox{90}{\hspace{0.0cm} $dR/dQ\rightarrow$kg/(y keVee)}
\includegraphics[height=.17\textheight]{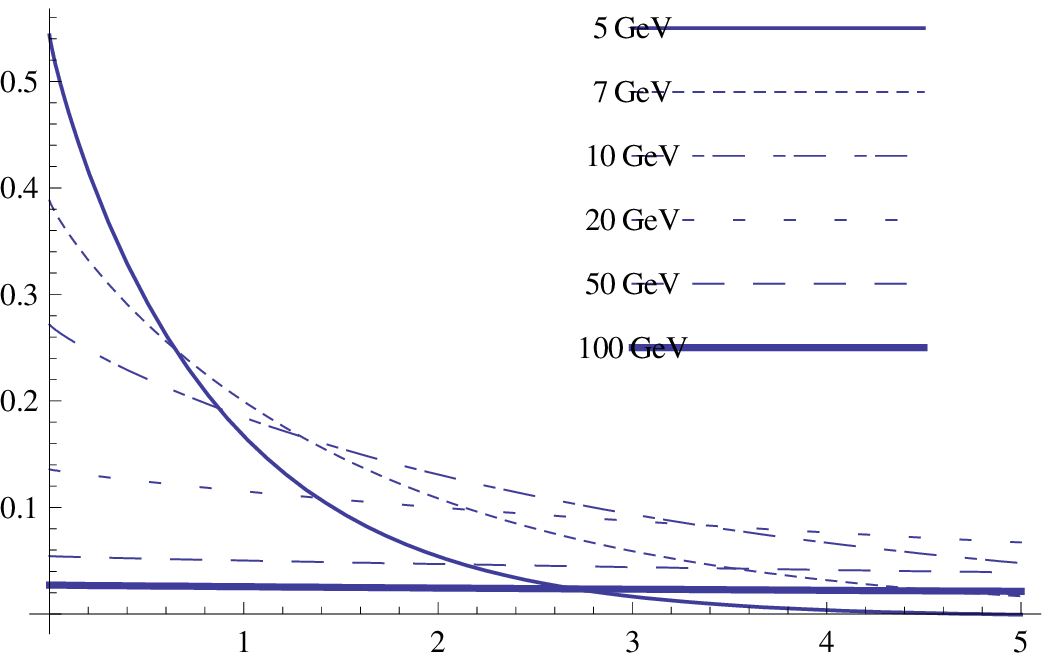}
}
\subfloat[]
{
\rotatebox{90}{\hspace{0.0cm} $d{\tilde H}/dQ\rightarrow$kg/(y keVee)}
\includegraphics[height=.17\textheight]{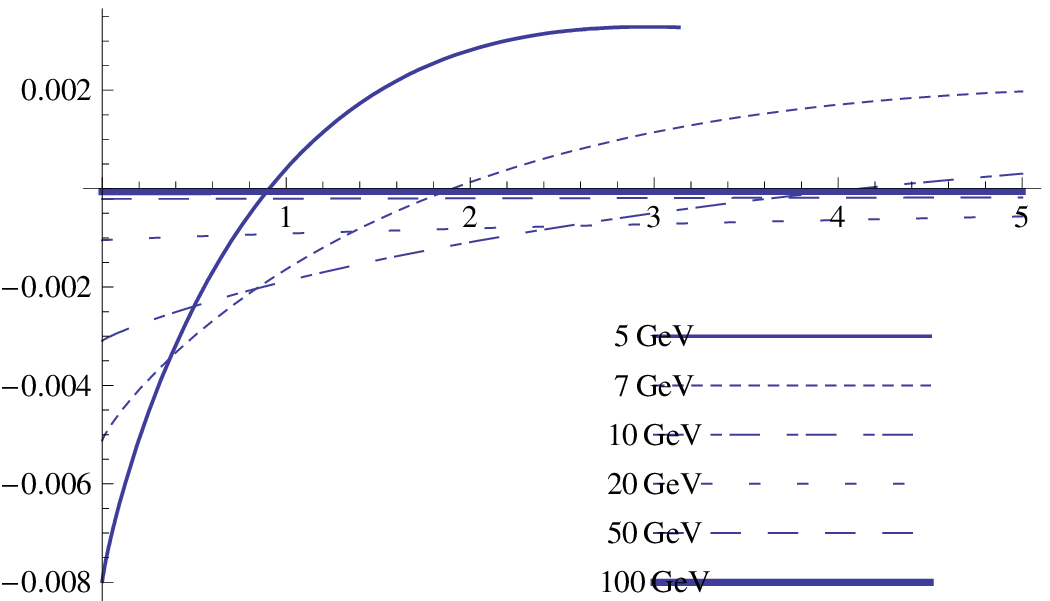}
}
\\
{\hspace{-2.0cm} $Q\rightarrow$keVee}
\caption{ The same as in Fig. \ref{fig:dRdQdHdQ_127} for the target $^{23}$Na.
 \label{fig:dRdQdHdQ_23}}
\end{center}
\end{figure}

We will apply the above formalism in the case of NaI, a target used in the DAMA experiment \cite{DAMA1,DAMA11}. The results for the Xe target are similar \cite{XENON10}. 
The differential rates $\frac{dR}{dQ}|_A$ and  $\frac{d\tilde{H}}{dQ}|_A$, for each component ($A=127$ and $A=23$) are exhibited in Fig. \ref{fig:dRdQdHdQ_127}-\ref{fig:dRdQdHdQ_23}. Following the practice of the DAMA experiment we express the energy transfer is in keVee using the phenomenological quenching factor \cite{LIDHART,SIMON03}. The nuclear form factor has been included (for a heavy target, like  $^{127}$I or $^{131}$Xe, its effect is sizable even for an energy transfer\cite{JDV12a} of 10 keV).
The  differential rate for the spin mode for low energy transfers is similar to those exhibited in Figs \ref{fig:dRdQdHdQ_127}-\ref{fig:dRdQdHdQ_23}, since the spin form factors are similar. They are, of course, simply scaled down by $A^2$, if one takes the spin cross section, a combination of the nuclear spin ME and the nucleon spin amplitudes, to be the same with the coherent nucleon cross section, i.e.  $\sigma^{\mbox{{\tiny spin}}}_{\mbox{{\tiny nuclear}}}=10^{-7}$pb. For the actual spin nucleon cross sections extracted from experiment see \cite{JDVSpin09} and \cite{PICASSO09,COUPP11,SIMPLE11}.
 
\begin{figure}
\begin{center}
\subfloat[]
{
\rotatebox{90}{\hspace{0.0cm} $H(a \sqrt{u}) \cos{\alpha}\rightarrow$}
\includegraphics[height=.15\textheight]{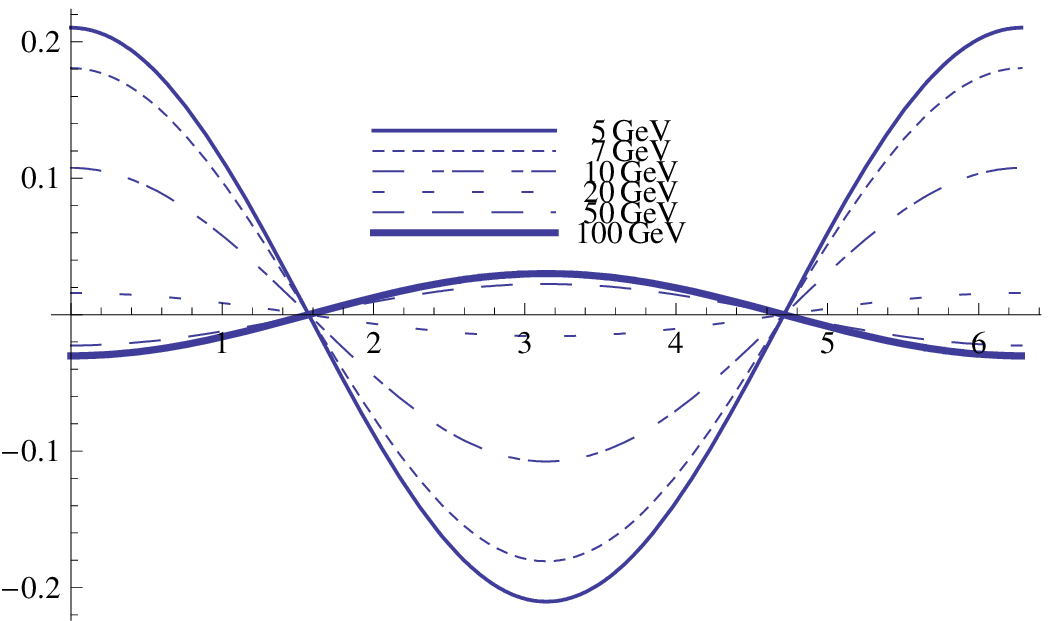}
}
\subfloat[]
{
\rotatebox{90}{\hspace{0.0cm} $H(a \sqrt{u}) \cos{\alpha}\rightarrow$}
\includegraphics[height=.15\textheight]{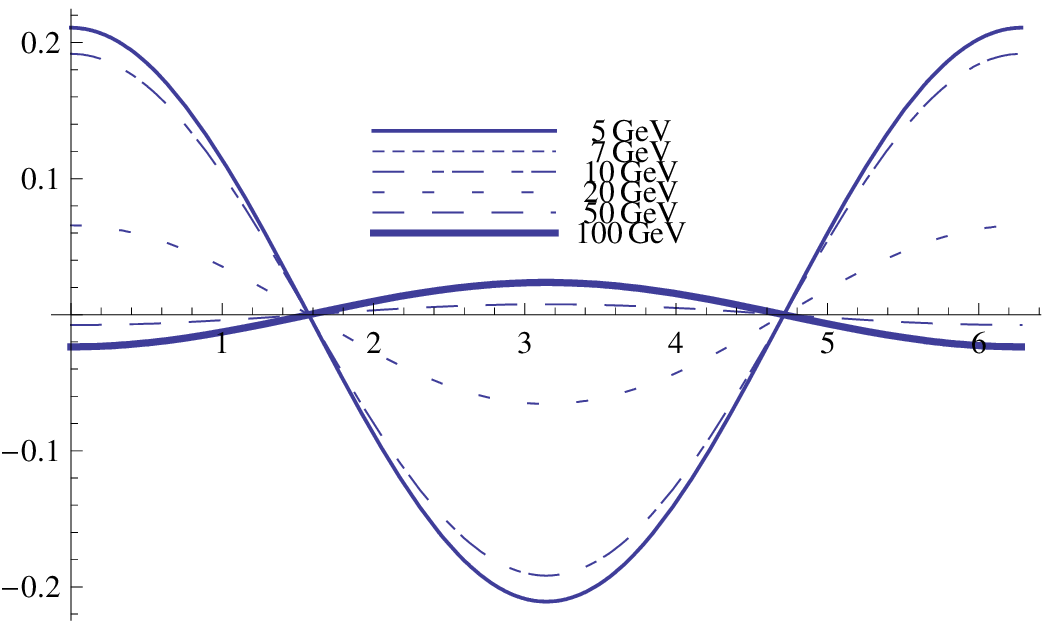}
}
\\
{\hspace{-2.0cm} $\alpha \rightarrow$}
\caption{ The modulation $H(a \sqrt{u}) \cos{\alpha}$ with an energy transfer of 2 keVee (a) and 5 keVee (b) for a heavy target (I or Xe). The solid, dotted, dot-dashed, dashed, long dashed and thick solid lines correspond to 5, 7, 10, 20, 50 and 100 GeV WIMP masses. Note that for some wimp masses on June 2nd the amplitude becomes negative (location of minimum rate). Note that the modulation is given relative to the time averaged rate.
 \label{fig:Hcosa127a}}
\end{center}
\end{figure}

\begin{figure}
\begin{center}
\subfloat[]
{
\rotatebox{90}{\hspace{0.0cm} $H(a \sqrt{u}) \cos{\alpha}\rightarrow$}
\includegraphics[height=.15\textheight]{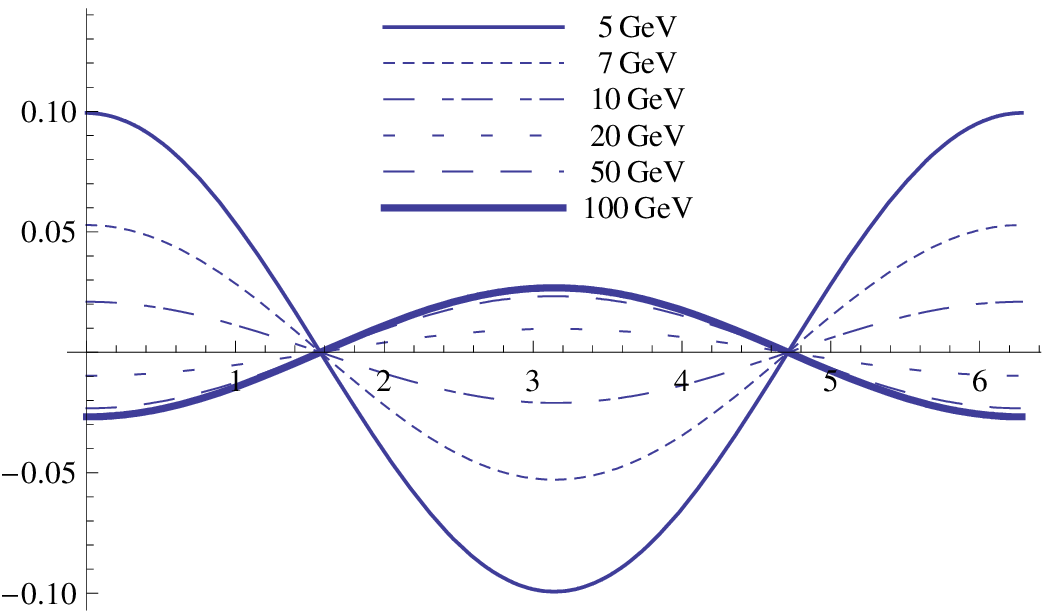}
}
\subfloat[]
{
\rotatebox{90}{\hspace{0.0cm} $H(a \sqrt{u}) \cos{\alpha}\rightarrow$}
\includegraphics[height=.15\textheight]{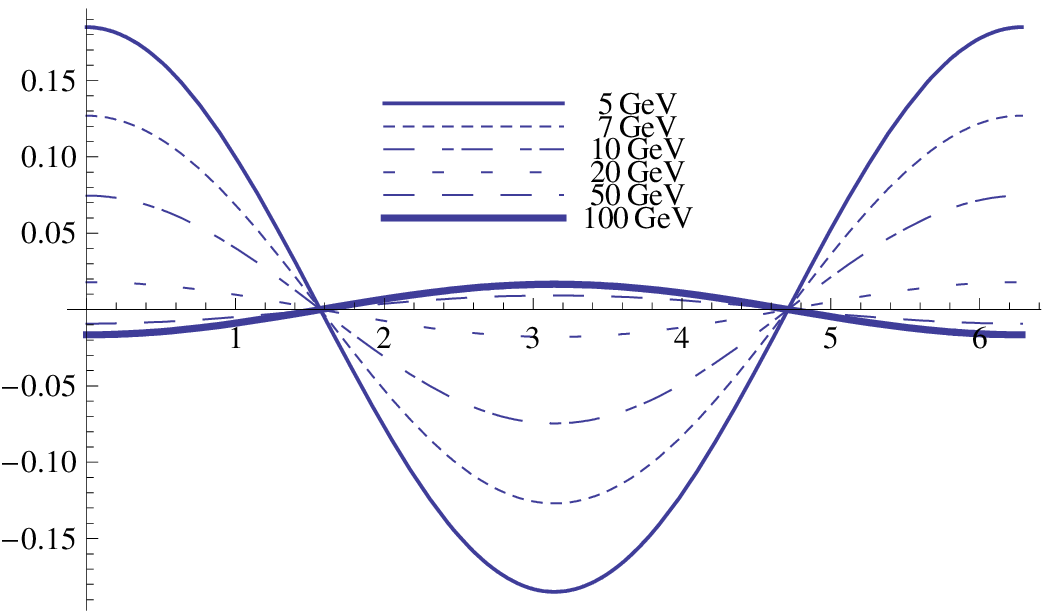}
}
\\
{\hspace{-2.0cm} $\alpha \rightarrow$}
\caption{ The same as in Fig. \ref{fig:Hcosa127a} for a light target (Na or F).
 \label{fig:Hcosa23a}}
\end{center}
\end{figure}

 The functions  $H(a \sqrt{u})\cos{\alpha}$  for each target component are shown in Figs  \ref{fig:Hcosa127a}- \ref{fig:Hcosa23a} as a function of $\alpha$ for various low energy transfers. The corresponding quantities for the spin mode are almost identical. We see that for certain values of the WIMP mass the modulation amplitude changes sign. This may perhaps by exploited to extract information on the WIMP mass from the data. A similar behavior has been found by considering various halo models and different minimum WIMP velocities \cite{GREEN04,SFG06}.
Sometimes, as is the case for the DAMA experiment, the target has many components. In such cases the above formalism can be applied as follows:
\beq
\frac{dR}{dQ}|_A\rightarrow \sum_i X_i\frac{dR}{dQ}|_{A_i},\quad u\rightarrow u_i,\quad X_i=\mbox{the fraction of the component } A_i\mbox{ in the target}
\eeq
\begin{figure}
\begin{center}
\subfloat[]
{
\rotatebox{90}{\hspace{0.0cm} $H(a \sqrt{u}) \cos{\alpha}\rightarrow$}
\includegraphics[height=.15\textheight]{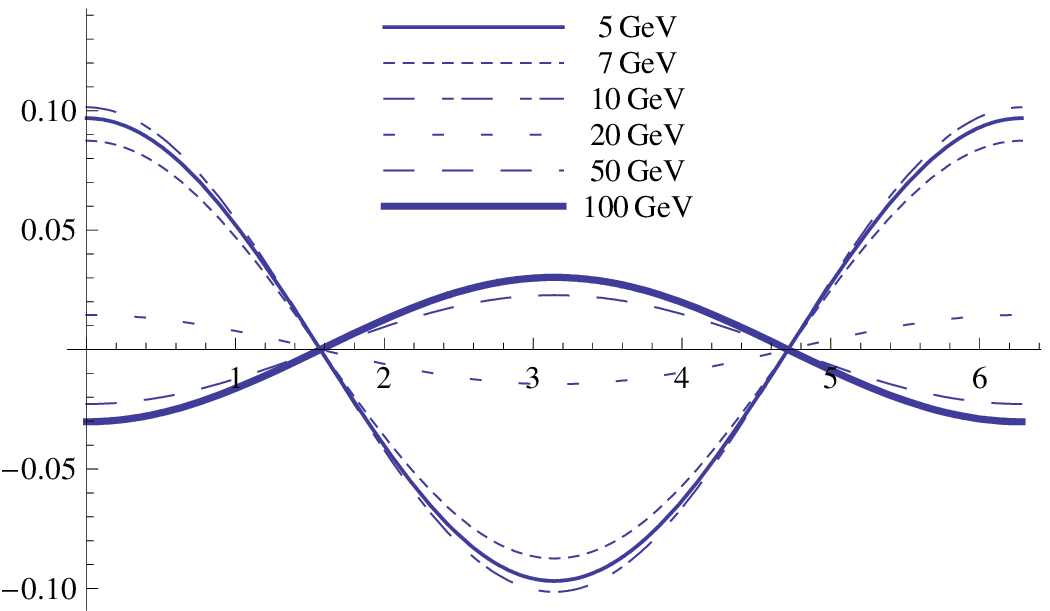}
}
\subfloat[]
{
\rotatebox{90}{\hspace{0.0cm} $H(a \sqrt{u}) \cos{\alpha}\rightarrow$}
\includegraphics[height=.15\textheight]{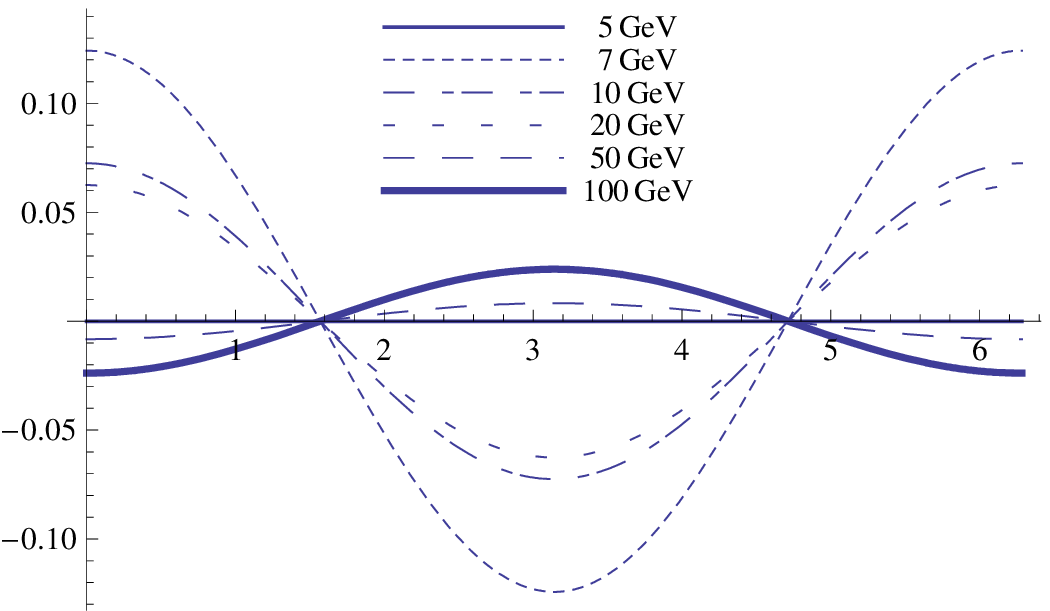}
}
\\
{\hspace{-2.0cm} $\alpha \rightarrow$}
\caption{ The same as in Fig. \ref{fig:Hcosa127a} for a NaI target.
 \label{fig:Hcosabotha}}
\end{center}
\end{figure}
The function $H(a\sqrt{u})\cos{\alpha}$ for NaI as a function of $\alpha$ is exhibited in Fig. \ref{fig:Hcosabotha}.
\section{Some results on total rates}

For completeness and comparison we will briefly present our results on the total rates. Integrating the differential rates discussed in the previous section we obtain the total time averaged rate $R_0$, the total modulated rate $\tilde{H}$ and the relative modulation amplitude$h$  given by:
\beq
R=R_0+\tilde{H} \cos{\alpha}\mbox{ or } 
R=R_0\left (1+h \cos{\alpha}\right )
\eeq
 Some special results in the case of low WIMP mass are exhibited in Tables \ref{tab1}-\ref{tab2}. From table \ref{tab2} it becomes clear that, for low mass WIMPs, large nucleon cross sections can accommodate the data. A similar interpretation holds for the \cite{CoGeNT11} data. In the case of non zero threshold one notices the strong dependence of the rime averaged rate on the WIMP mass. Also in this case the relative modulation $h$ substantially increases, the difference between the maximum and the minimum can reach 20$\%$. This however occurs at the expense of the number of counts, since both the time averaged and the time dependent part decrease, but the time averaged part decreases faster. So their ratio increases. This can be understood by noticing that the cancellation of the negative and positive parts in the differential modulated amplitide, becomes less effective in this case.

\begin{table}
\begin{center}
{\tiny
  \caption{Some  total event rates for some special WIMP masses and energy thresholds. The coherent nucleon cross section of $\sigma_n=10^{-7}$pb was employed.}
  \label{tab1}
  \begin{tabular}{|l|l|l|l|l|l|l|l|l|l|l|}
  \hline\hline
  & & & & & & & & & &\\
  \tiny
   $E_{th}$&$m_{\mbox{\tiny{WIMP}}}$  & $R_0$(I) & $\tilde{H}$(I) &h(I)& $R_0$(Na) & $\tilde{H}$(Na) &h(Na)& $R_0$(NaI) & $\tilde{H}$(NaI) &h(NaI)\\
       (keVee)& GeV &kg-y  &kg-y  & & kg-y & kg-y & &kg-y  & kg-y &\\
   \hline
 0 & 80 & 16.3 & -0.311 & -0.019 & 1.518 &
   0.028& 0.019 & 14.0 & -0.259& -0.018 \\
 0 & 20 & 25.8 & 0.285 & 0.011 & 2.35 & 0.050
   & 0.021& 22.2 & 0.249 & 0.019 \\
 0 & 10 & 18.4 & 0.356 & 0.019 & 2.045 & 0.046
   & 0.022& 15.9 & 0.309 & 0.019 \\
   \hline
 5 & 80 & 7.00 & -0.042 & -0.006 & 1.133 &
   0.038 & 0.034 & 6.11 & -0.030 & -0.005
   \\
 5 & 20 & 2.72 & 0.247 & 0.091 & 1.07 & 0.065
   & 0.060 & 2.47 & 0.219& 0.089 \\
 5 & 10 & 0.008 & 0.001 & 0.187 & 0.303 &
   0.031 & 0.103 & 0.053 & 0.006 & 0.114\\
   \hline
   \hline
   \end{tabular}
   }
\end{center}
\end{table}
\begin{table}
\begin{center}
{\tiny
  \caption{The same as in table \ref{tab1} for $\sigma_n=2 \times 10^{-4}$pb relevant for the DAMA region. One sees that, for very low mass WIMPs, large nucleon cross sections are required to obtain the rates claimed by the DAMA experiment \cite{DAMA11}.}
  \label{tab2}
  \begin{tabular}{|l|l|l|l|l|l|l|l|l|l|l|}
  \hline\hline
  & & & & & & & & & &\\
   $E_{th}$&$m_{\mbox{\tiny{WIMP}}}$  & $R_0$(I) & $\tilde{H}$(I) &h(I)& $R_0$(Na) & $\tilde{H}$(Na) &h(Na)& $R_0$(NaI) & $\tilde{H}$(NaI) &h(NaI)\\
       (keVee)& GeV &kg-y  &kg-y  & & kg-y & kg-y & &kg-y  & kg-y &\\
   \hline
 0 & 80 & $4.07\times 10^{4}$ & -776 & -0.019& $3.80\times 10^{3}$  & 70.2 &
   0.019 & $3.50\times 10^{4}$  & -647 & -0.018 \\
 0 & 20 & $6.43\times 10^{4}$  & 712 & 0.011 &$5.87\times 10^{3}$& 126 &
   0.021 & $5.54\times 10^{4}$ & 622 & 0.011 \\
 0 & 10 &$4.61\times 10^{4}$  & 891& 0.019 & $5.11\times 10^{3}$  & 115 &
   0.022 &$3.98\times 10^{4}$  & 772 & 0.019 \\
 5 & 80 & $1.75\times 10^{4}$  & -105 & -0.006 & $4.83\times 10^{3}$  & 95.0 &
   0.034 & $1.53\times 10^{4}$  & -74.6 & -0.005 \\
 5 & 20 & $6.80\times 10^{3}$ & 617 & 0.091& $2.69\times 10^{3}$  & 162 &
   0.060 & $6.17\times 10^{3}$  & 547 & 0.089 \\
 5 & 10 & 19.4 & 3.62 & 0.187 & 757 & 78.1 &
   0.103& 132 & 15.0 & 0.114\\
   \hline
   \hline
   \end{tabular}
   }
\end{center}
\end{table}
\section{Discussion}
In the present paper we obtained results on the differential event rates, both modulated and time averaged, focusing our attention on small energy transfers  and relatively light WIMPS. We found that:
\begin{itemize}
\item The relative modulation amplitude crucially depends on the WIMP mass. For small masses it exhibits normal behavior, but for large masses it changes sign (minimum in June). This effect is more pronounced in the case of heavy targets. We thus suggest an analysis of the experiments, in particular the DAMA experiment, along these lines to establish the location of the maximum on the $\alpha$-axis.
\item The relative modulation amplitude depends somewhat on the energy transfer, especially at low transfers.
\item For WIMP masses less than 10 GeV, the difference between the maximum and the minimum could reach between $20\%$ and $40\%$ for a heavy target, but it is a bit less for a light target, depending on the enegy transfer.
\item The relative modulation amplitude for NaI is the weighted average of its two components, and in the low energy regime, between 1 and 6 keVee, it does not change much with the energy transfer.
\item Once it is established that one actually observes the modulation effect, the sign of the modulation may be exploited to infer the WIMP mass.
\end{itemize}
For low WIMP mass  the total rates depend strongly on the threshold energy, especially for a heavy target. The relative modulation in the presence a threshold gets quite large ($h\approx 0.2$), but, unfortunately, this occurs at the expense of the number of counts. It is important to compare the relative total modulation in a least one light and one heavy target. For very low energy thresholds, if the signs are opposite, one may infer that the WIMP is heavy, $m_{\mbox{\tiny{WIMP}}}\ge100$ GeV.
\section*{Acknowledgments} The author is indebted to the CERN Theory Division and KITPC/ITP-CAS for their hospitality and support. This work was partially supported by UNILHC PITN-GA-2009-237920.
\section*{References}

\end{document}